\documentclass[authoryear,review]{elsarticle}

\pdfoutput=1

\usepackage{amsmath}
\usepackage{graphicx}

\journal{}

\begin{document}
	
	\begin{frontmatter}
		
		
		\title{Assessment of Sea-Level Rise Impacts on Salt-Wedge Intrusion in Idealized and Neretva River Estuary}
		
		\author[gradri]{Nino Krvavica\corref{mycorrespondingauthor}}
		\cortext[mycorrespondingauthor]{Corresponding author}
		\ead{nino.krvavica@uniri.hr}
		
		\author[gradri]{Igor Ru\v{z}i\'{c}}
		
		\address[gradri]{University of Rijeka, Faculty of Civil Engineering, Radmile Matejcic 3, 51000 Rijeka, Croatia}

		\begin{abstract}
			
		Understanding the response of estuaries to sea-level rise is crucial in developing a suitable mitigation and climate change adaptation strategy. This study investigates the impacts of rising sea levels on salinity intrusion in salt-wedge estuaries. The sea-level rise impacts are assessed in idealized estuaries using simple expressions derived from a two-layer hydraulic theory, and in the Neretva River Estuary in Croatia using a two-layer time-dependent model. The assessment is based on three indicators - the salt-wedge intrusion length, the seawater volume, and the river inflows needed to restore the baseline intrusion. The potential SLR was found to increase all three considered indicators. Theoretical analysis in idealized estuaries suggests that shallower estuaries are more sensitive to SLR. Numerical results for the Neretva River Estuary showed that SLR may increase salt-wedge intrusion length, volume, and corrective river inflow. However, the results are highly non-linear because of the channel geometry, especially for lower river inflows. A theoretical assessment of channel bed slope impacts on limiting a potential intrusion is therefore additionally discussed. This findings emphasize the need to use several different indicators when assessing SLR impacts. 
		
		\end{abstract}
		
		\begin{keyword}
			climate change \sep sea level changes \sep salt-wedge estuaries \sep salinity stratification \sep two-layer models \sep saline intrusion
			
		\end{keyword}

	\end{frontmatter}

\newpage

\section{Introduction}

One of the most prominent consequences of ongoing climate change is the sea-level rise (SLR). The most recent report on changes in global and regional mean sea levels, sea-level extremes and waves is provided in the Fifth Assessment Report of the Intergovernmental Panel for Climate Change (IPCC) \citep{church2013}. Although the effects that climate change will have in the future are uncertain, all recent studies agree that global sea levels will rise to some extent. A global SLR of 19 cm has been recorded over the period 1901–2010, with long-term projections suggesting that an increase with respect to 1986-2005 mean will lie within the range of 32–82 cm by the end of the twenty-first century \citep{church2013}.

Sea-level rise has been recognized as a major threat to low-lying coastal environments, such as wetlands, salt marshes, and lagoons, with potentially devastating consequences \citep{kirwan2010,cui2015vulnerability,carrasco2016,sampath2016,bigalbal2018}. 
Estuaries are especially vulnerable coastal areas, because the seawater intrusion may endanger a fragile estuarine ecosystem and decrease the freshwater quality used for agricultural or water supply systems. Therefore, predicting and quantifying the salinity changes along estuaries is one of the key issues for water management in coastal areas.

In the past decade, numerous studies have investigated the impacts of rising sea levels in estuaries. Majority of these studies have found that SLR changes the water salinity in estuaries. In particular, increased seawater intrusion lengths and higher average salinities resulting from SLR have been predicted in the San Francisco Bay \citep{chua2011sensitivity,chua2014impacts}, Delaware Estuary \citep{ross2015}, Snohomish River Estuary \citep{yang2015estuarine}, Chesapeake Bay \citep{hong2012responses} and its tributaries James and Chickahominy Rivers \citep{rice2012} in USA, Yangtze River Estuary \citep{li2015water,chen2016influence} and Nandu River Estuary \citep{he2018effect} in China, Simjin River Estuary in Korea \citep{shaha2012}, Wu River Estuary in Taiwan \citep{chen2015modeling}, Gorai River network in Bangladesh \citep{bhuiyan2012}, Bahmanshir Estuary in the Persian Gulf \citep{etemad2015}, Sebua Estuary in Morocco \citep{haddout2018}, Gironde Estuary in France \citep{vanmaanen2018}, as well as 96 estuaries in England and Wales \citep{prandle2015}.

\cite{tian2019factors} examined several factor that control the seawater intrusion and found that the river flow rate is the dominant factor, followed by the sea level, which prevails in seasonal variations. Wind, on the other hand, has an impact only in shorter time scales. The channel geometry is also an important factor. Namely, the channel meanders can reduce stratification, change the two-layer circulation, and therefore alter the potential seawater intrusion \citep{tian2019factors}. In addition to increasing seawater intrusion and average salinity, several recent studies have found that SLR may also strengthen stratification \citep{chua2011sensitivity}, weaken vertical exchange \citep{hong2012responses}, increase seawater volume \citep{hong2012responses}, increase residence time and water age \citep{hong2012responses, chen2015modeling}, and cause tidal amplifications \citep{kuang2014,vanmaanen2018}. 

Almost all studies mentioned above have focused on well-mixed or partially-mixed estuaries, which are characterized by homogeneous vertical column or week stratification and gradual salinity changes along an estuary. On the other hand, studies investigating SLR impacts in salt-wedge estuaries, where freshwater and seawater layers are separated by a sharp interface (pycnocline), are less common. Strong vertical stratification, however, is found in many estuaries worldwide. Some prominent examples of salt-wedge estuaries are Mississippi and Merrimack Rivers in USA \citep{geyer2014estuarine}, Fraser River in Canada \citep{macdonald2004turbulent}, or the Ebro and Rhone Rivers in Europe \citep{ibanez1997}.
Strong stratification in estuaries is maintained by high river inflows, which dampen the vertical mixing caused by tidal motions \citep{geyer2014estuarine}. 
In micro-tidal conditions, stratification is usually persistent under all river inflows \citep{geyer2011}. In the Mediterranean region, for example, estuaries are strongly stratified even for relatively low river inflows \citep{ibanez1997,ljubenkov2012,krvavica2016field,krvavica2016salt}. 
On the other hand, in tidally more dynamic environments, such as the Fraser Estuary, a strongly stratified structure may also be observed, but only during higher inflows \citep{macdonald2004turbulent}.

Although the dynamics of salt-wedge estuaries is well documented \citep{geyer2011}, there are hardly any studies available on the impact of SLR on the salt-wedge intrusion. This critical issue has been partially investigated in the Nile Estuary in Egypt \citep{mahgoub2014} and Rje\v{c}ina River Estuary in Croatia \citep{krvavica2016salt}. Both these studies showed that the tip of the salt-wedge may intrude further upstream under potential SLR. Recently, \cite{leung2018} simulated the salt-wedge behaviour in the Fraser River Estuary in Canada under several climate and man-made changes. They also reported increased intrusion lengths, but additionally found that during low flow conditions, SLR may become more significant than changes in the river flow regime. 

Considering that there are only a few comprehensive studies on SLR impacts on the salt-wedge intrusion, and almost no such studies in microtidal environments, the main aim of this study is to expand the knowledge on this topic.
In contrast to previous studies that presented their findings for individual estuaries, this work is based on a more theoretical analysis of idealized estuaries, and numerical modelling of a salt-wedge estuary characteristic for microtidal environments. The theoretical analysis and simple analytical expressions are derived from a two-layer hydraulic theory, which is considered a satisfactory approximation in strongly stratified conditions. A wide range of idealized estuaries is considered here to provide a more general assessment of SLR impacts on seawater intrusion, and to remove the influence of channel geometry, in particular, bed elevations and width variations. The findings from theoretical analyses are substantiated by an example of a microtidal Neretva River Estuary in Croatia. Furthermore, along with the seawater intrusion length, as a common indicator for SLR impacts, we propose two additional indicators - the seawater volume and the river inflow increase needed to restore the baseline intrusion. 

The proposed aims arise from two main hypothesis of this work: a) theoretical analysis derived from a two-layer hydraulic theory is an adequate tool for a generalized assessment of the SLR impacts on seawater intrusion in salt-wedge estuaries, and b) SLR impacts can be quantified by computing the changes in the salt-wedge intrusion length and volume, as well as the river inflow increase needed to restore the baseline intrusion.

\section{Study area}

\subsubsection{Neretva River}

Neretva River is the largest river in the eastern part of the Adriatic Basin. It is 215 km long, and for the most part, it flows through Bosnia and Herzegovina (BiH). For the final 22 km, it flows through Croatia before it enters the Adriatic Sea. Its catchment area is estimated to 10,240 km$^2$, of which only 280 km$^2$ is located in Croatia \citep{ljubenkov2012,oskorus2019}. Neretva River is a typical mountain river in its upper and middle part, but in the lower reaches, it forms a wide alluvial delta covering approximately 12,000 hectares (Fig.~\ref{fig:Neretva_geom}).

\subsubsection{Neretva River Estuary}

Lower reaches of the Neretva River are one of the most important and productive agricultural areas in Croatia. Unfortunately, the vicinity of the Adriatic Sea poses a threat to the future of the agricultural industry in this area. Namely, there are significant challenges with deteriorating freshwater quality for irrigation caused by the salinity intrusion, as well as persistent salinization of the groundwater and soil during summer months. These problems are expected to become even more severe in the future under climate change and a potential rise of sea levels, as well as the reduction of freshwater inflow during the summer.

\begin{figure}
	\centering
	\includegraphics[width=12 cm]{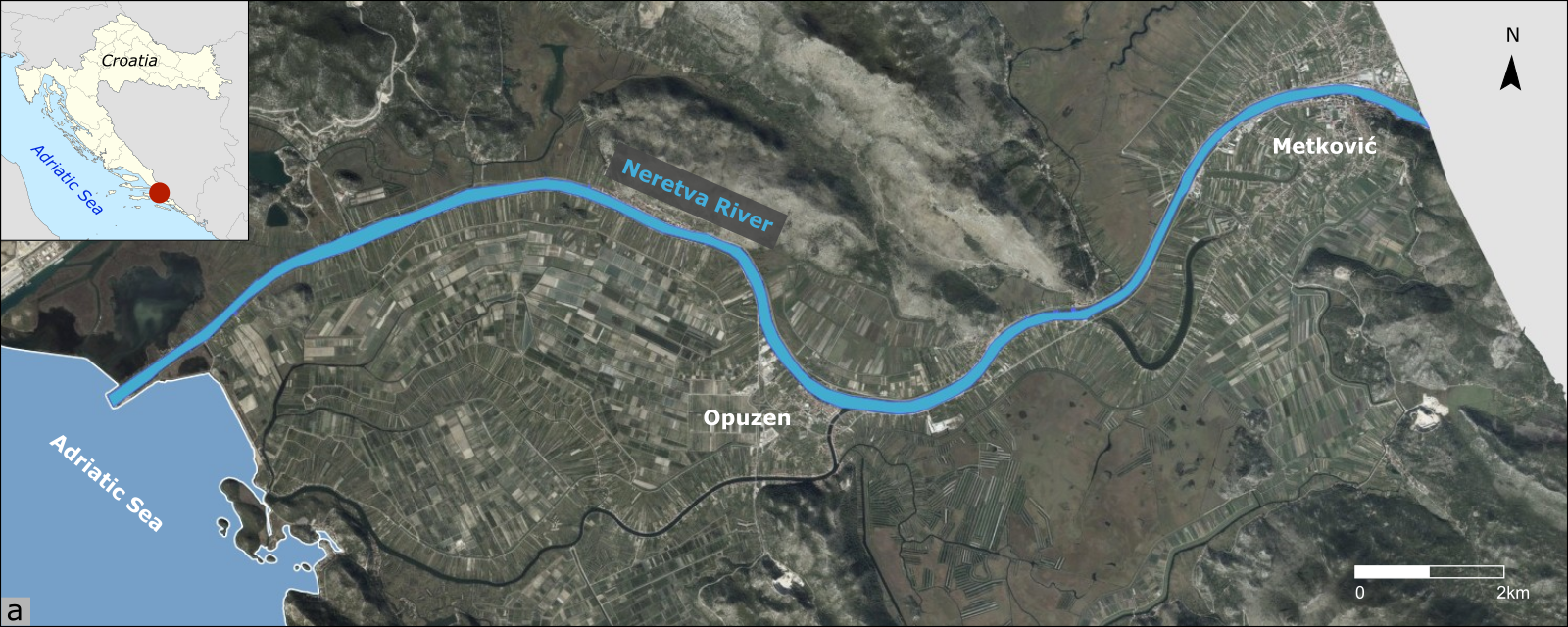}
	\vfill
	\includegraphics[width=6 cm]{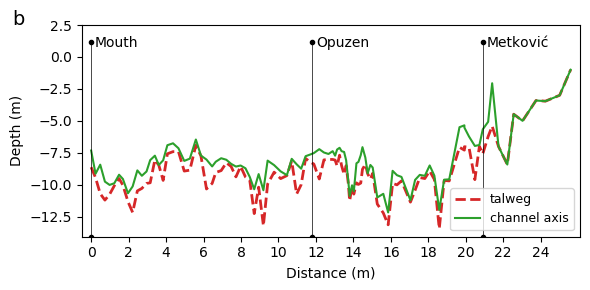}
	\hfill
	\includegraphics[width=6 cm]{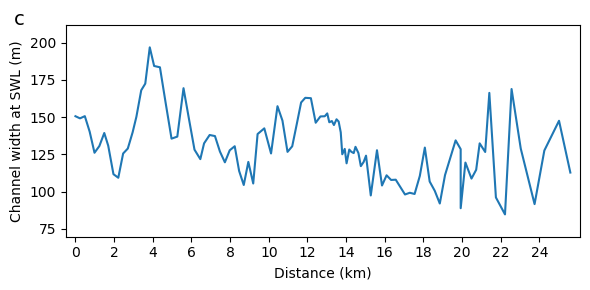}
	\caption{Neretva River Estuary: a) geographical location and digital ortophoto, b) longitudinal bottom profile and c) channel width variations at SWL ($\pm 0$ m a.s.l.)}
	\label{fig:Neretva_geom}
\end{figure}

The location and channel geometry of the Neretva River Estuary is shown in Fig.~\ref{fig:Neretva_geom}. Its mean channel depth is 8.5 m and the mean channel width is 132 m, both with some variations. 
The dynamics of the Neretva River Estuary is governed by tidal oscillations of the Adriatic Sea and upstream inflows, which depend on the operational aspects and management of the hydro-power system of accumulations and dams built in its middle reaches located in BiH. 
Neretva River has a pronounced seasonal character, typical for Mediterranean climate, with a well defined high flow season from October to April, and a low flow period from May to September. 
During the summer, when irrigation needs are more pronounced, the precipitation and river inflows are at their lowest, and the seawater regularly intrudes over 20 km upstream \citep{ljubenkov2012}.

Although water levels in its lower part have been measured since 1957, a systematic flow rate and conductivity measurements were established only in 2015. Short-term data at the station Metković indicates that its mean annual flow rate is 325 m$^3$/s, while a maximum measured flow rate is 1375 m$^3$/s \citep{oskorus2019}. Minimum flow rate is regulated by an international agreement between Croatia and BiH, which states that a minimum 50 m$^3$/s must be released from the hydro-power plant Mostar \citep{oskorus2019}.

\subsubsection{Tides and sea levels in the Adriatic Sea}

The tides in the Adriatic Sea are of a mixed type with relatively low amplitudes. The maximum astronomical tidal amplitude recorded between 1953 and 2006 at the Bakar tidal gauging station was 123 cm; however, mean tidal amplitudes do not exceed 30 cm \citep{rezo2014}. Highest recorded sea level at the Neretva River mouth (in period 1977-2018) was 120 cm a.s.l., and the lowest -25 cm a.s.l. \citep{oskorus2019}.

Because of the microtidal environment and high freshwater flow rates, the Neretva River Estuary is a typical salt-wedge estuary according to classification by \cite{geyer2014estuarine}, occupying the upper-left corner of the estuarine parameter space with freshwater Froude number higher than 0.1 and mixing number under 0.4.
Salinity profile measurements conducted by \cite{ljubenkov2012} shown very strong stratification with pycnocline thickness always under 0.5 m. Similar salinity profiles in the Rje\v{c}ina River (also located on the eastern Adriatic coast) indicated buoyancy frequencies well over $N^2>0.1$ \citep{krvavica2016salt}. \cite{geyer2008quantifying} reported typical values for $N^2$ ranging from 0.0025 to 0.01 $s^{-2}$ for partially mixed estuaries and up to 0.1 $s^{-2}$ for salt-wedge estuaries. In this case, the stratification is so pronounced that it even exceeds the upper values usually associated with salt-wedge estuaries. Such a strong stratification is a result of extremely low tidal dynamics of the Adriatic Sea.

A comprehensive overview of the literature on the sea-level research in the Adriatic Sea is given by \cite{vilibic2017}. The Mediterranean Sea, and the Adriatic Sea as its integral part, has been recognized as one of the most sensitive to climate change. Some projections for the period between 1990-2000 and 2040-2050 show up to 25.6 cm increase averaged over the Mediterranean (and Adriatic) Sea. Projections for the basin-averaged steric SLR in the Adriatic Sea over the twenty-first century point to only 13 cm. However, when the contribution of salinity and temperature are taken into account to asses both mass and steric components on a local scale, a sea-level rise of 49.7 cm is expected for the period 2000-2100. For a more detailed explanation of the salinity and temperature contribution to SLR in the Mediterranean basin see the study by \cite{jorda2013interpretation}. 
This is close to findings by \cite{tsimplis2012}, who used tidal gauge records from almost all stations in the Adriatic Sea to analyse trends in sea levels and estimated an increase in the range from $2.0 \pm 0.9$ to $3.4 \pm 1.1$ mm/year, which suggests that SLR by the year 2100 could reach up to 45 cm.
More recent analysis, however, suggests that the sea-level rise is rapidly increasing \citep{church2013}, and that trends derived from past and present measurements underestimate potential SLR in the future. Therefore, we considered a more extreme scenario in the Adriatic Sea with sea levels rising to 100 cm with respect to present conditions.

\section{Materials and Methods}

\subsection{Hydraulics of Salt-Wedge Estuaries}

\begin{figure}
	\centering
	\includegraphics[width=9 cm]{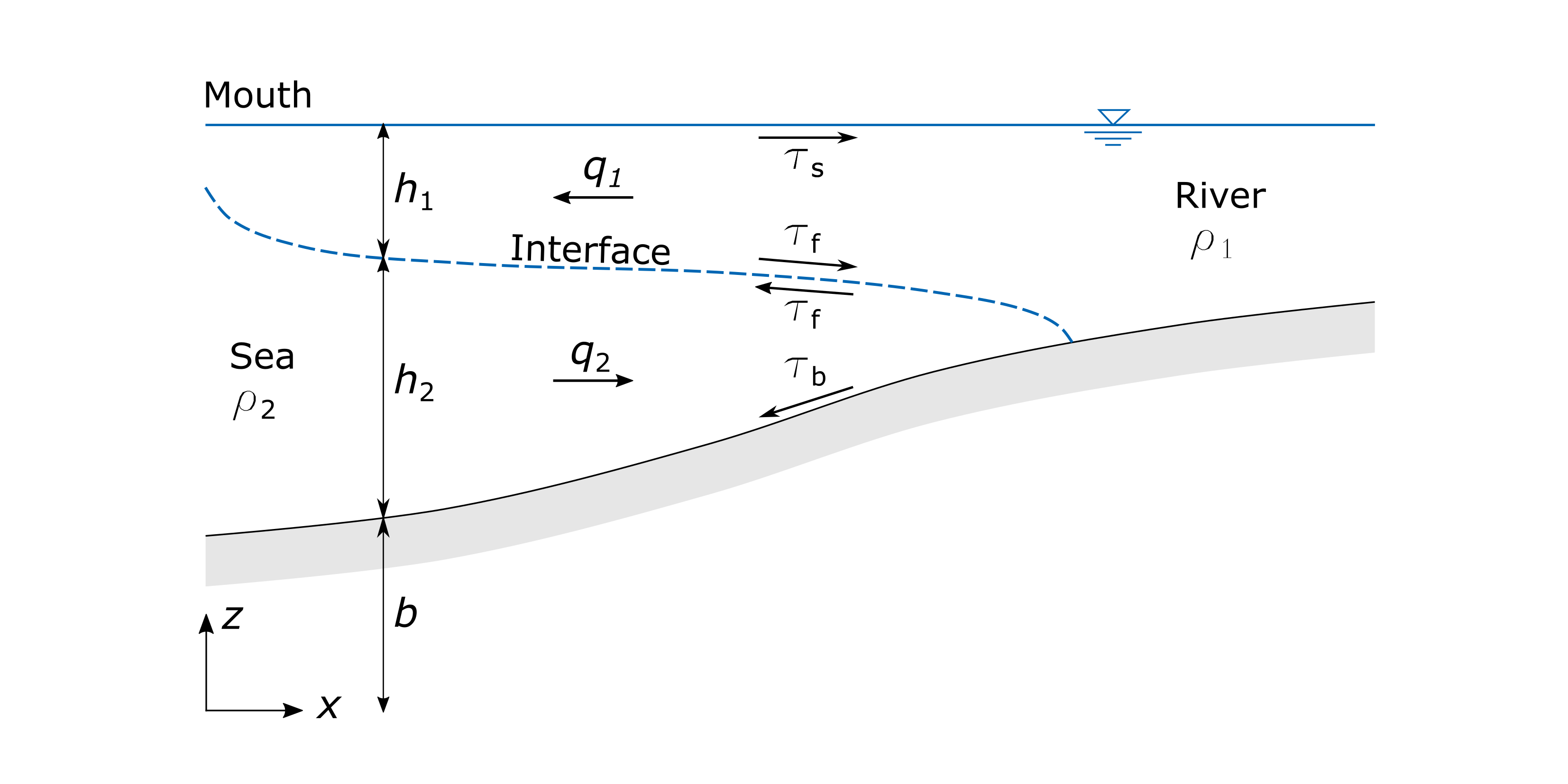}
	\caption{Characteristic longitudinal section of a salt-wedge estuary}
	\label{fig:longitudinal}
\end{figure}

Although three-dimensional (3D) hydrodynamic models are regularly used for computing physical process in estuaries \citep{mohammed2017}, salt-wedge estuaries are adequately represented by a simpler two-layer hydraulic theory \citep{sargent1987,krvavica2016salt,krvavica2017}. Two-layer theory is derived from coupled shallow water equations (SWE) - we assume that the vertical structure can be represented by two shallow layers of different densities separated by a pycnocline of zero thickness (see Fig.~\ref{fig:longitudinal}). Furthermore, we assume, under the usual SWE theory, that the velocity and density are uniform over each layer, vertical accelerations are negligible, the pressure is hydrostatic, the channel bed slope is mild, and the viscous effects such as friction and turbulence are represented by simple empirical equations. Also, we assume that the flow is predominately one-dimensional (1D) and that the channel is prismatic with rectangular cross section of uniform width.

\cite{schijf1953} and \cite{stommel1953} were among first to develop a mathematical theory for two-layer flows. Recently, \cite{geyer2011} provided a good insight into the current knowledge on the dynamics of strongly stratified estuaries.
The governing system of partial differential equations (PDEs) for a one-dimensional two-layer shallow water flow in prismatic channels with constant width (Fig. \ref{fig:longitudinal}) is written as a coupled system of continuity and momentum equations \citep{schijf1953}:
\begin{equation}
\frac{\partial h_1}{\partial t} + 
\frac{\partial q_1}{\partial x} = 0
\label{eq:CMass1}
\end{equation}
\begin{equation}
\frac{\partial q_1}{\partial t} +
\frac{\partial}{\partial x} \left( \frac{q_1^2}{h_1} + \frac{g}{2}h_1^2\right) = 
- gh_1\frac{\textrm{d}b}{\textrm{d}x} - 
gh_1\frac{\partial h_2}{\partial x} - 
\frac{\tau_f}{\rho_1}
\label{eq:CMom1}
\end{equation}
\begin{equation}
\frac{\partial h_2}{\partial t} + 
\frac{\partial q_2}{\partial x} = 0
\label{eq:CMass2}
\end{equation}
\begin{equation}
\frac{\partial q_2}{\partial t} +
\frac{\partial}{\partial x} \left( \frac{q_2^2}{h_2} + \frac{g}{2}h_2^2\right) = 
- gh_2\frac{\textrm{d}b}{\textrm{d}x} - 
gh_2\frac{\rho_1}{\rho_2}\frac{\partial h_1}{\partial x} +
\frac{\tau_f - \tau_b}{\rho_2}
\label{eq:CMom2}
\end{equation}
%
where $x$ is the coordinate along the channel axis, $t$ is time, $h_i(x,t)$ is the thickness of $i$-th layer, $b(x)$ is the channel bottom function, $q_i(x,t)$ is the layer flow rate per unit width, $\rho_i$ is the layer density, $g$ is acceleration of gravity, $\tau_f$ is the interfacial shear stress, $\tau_b$ is the shear stress between channel bed and lower layer fluid, and subscript $i=1,2$ denotes the upper and lower layer, respectively. 
The shear stresses may be parametrized using a quadratic drag equation:
\begin{equation}
\tau_f = C_f \rho_1 \vert u_1 - u_2 \vert \left( u_1 - u_2 \right)
\end{equation}
and
\begin{equation}
\tau_w = C_w \rho_2 \vert u_2 \vert u_2
\end{equation}
where $C_f$ is the interfacial friction factor, $C_b$ is the bed friction factor and $u_i(x,t) = q_i(x,t) / h_i(x,t)$ is the layer velocity. 

At constant sea-water level (SWL) and river inflow $q$, an equilibrium is established between the buoyancy pressure gradients, friction forces, and inertial forces, and eventually a steady-state flow forms \citep{sargent1987}. This two-layer stationary form is sometime called an arrested salt-wedge, which is characterized by a stagnant lower layer called a salt-wedge because of its shape, and an upper layer of constant flow rate. 

A mathematical model for an arrested salt-wedge is derived from Eqs.~(\ref{eq:CMass1})-(\ref{eq:CMom2}) following \cite{schijf1953}. Time derivatives are neglected, negative flow is asserted in the upper layer $q_1=q<0$ (because $x$ increases in the upstream direction) and zero-flow in the lower layer $q_2=0$. Momentum equations are then subtracted from each other under the rigid lid approximation, and a single ordinary differential equation (ODE) is obtained that describes the pycnocline depth along the channel axis:
\begin{equation}
\frac{\textrm{d}h_1}{\textrm{d}x} =
 \frac{F_d^2}{1 - F_d^2} C_f \left(1  +r \frac{h_1}{h_2}\right).
\label{eq:ode}
\end{equation}
where $r=\rho_1 / \rho_2$ is the density ratio and $F_d$ is the internal Froude number for the upper layer:
\begin{equation}
F_d = \frac{q_1}{\sqrt{g(1-r)h_1^3}}.
\label{eq:Fd}
\end{equation}

Note that Eq. (\ref{eq:ode}) corresponds to the one derived by \cite{geyer2011} or \cite{krvavica2014hydraulics} in a slightly different form. This equation defines how the upper layer changes along the estuary, but it is not valid upstream from the tip of the salt-wedge. Also, it is clear that the solution is singular when the internal Froude number is equal to one $F_d = 1$. This may occur at the hydraulic control, which is usually located at the river mouth, where the numerical integration should begin \citep{geyer2011}. Note that the hydraulic control may also be present at other locations along channels, if there are strong lateral constrictions or abrupt bed elevations \citep{poggioli2015}.

For idealized estuaries, Eq.~(\ref{eq:ode}) may be further simplified by considering a horizontal channel bed ($b(x)=\textrm{const.}$) and introducing the following non-dimensional parameters:
\begin{equation}
\varphi=h_1/H_0, \quad \chi=C_fx/H_0, 
\label{eq:HX}
\end{equation}
where $H_0$ is a constant total depth in a prismatic channel with no bed elevations. Also, the internal Froude number for the river \citep{geyer2011} is
\begin{equation}
F_0 = F(q, H_0, r) = \frac{q}{\sqrt{g(1-r) H_0^3}} = F_d \varphi^{3/2} .
\label{eq:F0}
\end{equation}
Note, that $F_0$ is the same as $F_d$ when the upper layer flow rate $q_1$ equals river inflow $q$ (which is always true for idealized arrested salt-wedges) and thickness $h_1$ equals total depth $H_0$ (which is true for idealized channels upstream from the tip of the salt-wedge). Furthermore, this non-dimensional parameter corresponds to the net barotropic flow $U_0$ introduced by \cite{armi1986} as the key dynamical parameter governing a two-layer hydraulic flow.

Substituting the parameters defined by Eqs.~(\ref{eq:HX}) and (\ref{eq:F0}), and using the Boussinesq approximation, which neglects the density ratio $r$ everywhere expect where multiplied by $g$ \citep{tritton1988}, Eq.~(\ref{eq:ode}) becomes
\begin{equation}
\frac{\textrm{d}\varphi}{\textrm{d}\chi}=
\frac{F_0^2}{\varphi^3-F_0^2} \left(\frac{1}{1-\varphi}\right).
\label{eq:odenondim}
\end{equation}

Equation (\ref{eq:odenondim}) is then integrated over non-dimensional depth to obtain an integral form of the equation:
\begin{equation}
\chi = \int_{\varphi_1}^{\varphi_2} \left[\frac{F_0^2}{\varphi^3-F_0^2} \left(\frac{1}{1-\varphi}\right)\right] ^{-1} \textrm{d}\varphi.
\label{eq:chi}
\end{equation}

To find the salt-wedge length, Eq.~(\ref{eq:chi}) is integrated between the hydraulic control (usually located at the river mouth) and the tip of the salt-wedge. These boundaries are introduced as the lower and upper integral limits.
Considering that $F_d=1$ is true at the hydraulic control point, we can combine Eq.~(\ref{eq:Fd}) and Eq.~(\ref{eq:F0}) to obtain non-dimensional upper and lower layer depths at the river mouth (critical depths):
\begin{equation}
\varphi_{cr} = h_{1,cr} / H_0 = F_0^{2/3}
\label{eq:h1cr}
\end{equation}
and
\begin{equation}
1 - \varphi_{cr} = h_{2,cr} / H_0 = 1 - F_0^{2/3}.
\label{eq:h2cr}
\end{equation}
Similarly, at the salt-wedge tip we have $h_1 = H_0$, therefore $\varphi_2 = 1$.

Integrating Eq.~(\ref{eq:chi}) between $\varphi_1 = F_0^{2/3}$ and $\varphi_2 = 1$ leads to the following analytical expression for a non-dimensional length $L^*$ of the salt-wedge intrusion: 
\begin{equation}
	L^*(F_0) = \frac{1}{20}
	\left(15F_0^{2/3} - 6 F_0^{4/3} -10 + F_0^{-2} \right).
	\label{eq:LH}
\end{equation}
A dimensional salt-wedge length $L$ can then be restored from Eq.~(\ref{eq:HX}) by computing $L=H_0L^*/C_f$, and then we obtain a well-known analytical expression derived by \cite{schijf1953}:
\begin{equation}
	L(H_0, C_f, F_0) = \frac{H_0}{20C_f}
	\left(15F_0^{2/3} - 6 F_0^{4/3} -10 + F_0^{-2} \right).
	\label{eq:L}
\end{equation}
It is clear that the intrusion length is a function of the channel depth $H_0$, interfacial friction factor $C_f$ and internal Froude number $F_0$ (defined upstream from the salt-wedge).

The shape of the salt-wedge may be computed similarly, but in a discrete form by repeatedly integrating Eq.~(\ref{eq:chi}), for a series of boundary values starting from $\varphi_1 = F_0^{2/3}$ and $\varphi_{2} = \varphi_1 + \Delta \varphi$, where $\Delta \varphi$ is a spatial step. We increase $\Delta \varphi$  until we reach $\varphi_{2} = 1$. In this way, each $\varphi_{2}$ is the non-dimensional upper layer thickness at a computed non-dimensional distance $\chi$ from the hydraulic control point. Discrete function $\varphi(\chi)$,  therefore, gives a non-dimensional function for the thickness of the salt-wedge, i.e. pycnocline depth.

We see that arrested salt-wedges are fully defined by three components - initial pycnocline depth, i.e. critical lower layer depth $h_{2,cr}$ or $1-\varphi_{cr}$, length $L$ or $L^*$, and pycnocline function, i.e., thickness depth variability from the salt-wedge tip to the hydraulic control point defined by set of points $(h_2, x)$ or  $(1-\varphi_2, \chi)$ in a non-dimensional form. 

Combining these parameters, we can derive an additional one - salt-wedge volume per unit width $V$, defined as the total volume of seawater below the pycnocline, which is obtained by integrating $h_2(x)$ from the river mouth ($x=0$) to the salt-wedge tip ($x=L$):
\begin{equation}
V(L, h_2) = \int_{0}^{L} h_2(x) \textrm{d}x,
\label{eq:V}
\end{equation}
or in a non-dimensional form:
\begin{equation}
V^*(L, \varphi) = \frac{V}{LH_0} = \frac{1}{L} \int_{0}^{L} (1-\varphi) \textrm{d}x.
\label{eq:Vnondim}
\end{equation}
Note that value $V^*$ represents the total seawater to fresh-water ratio along the salt-wedge. This information may sometimes be more important than the intrusion length as it directly relates to salinity intrusion into aquifer, especially in highly permeable mediums \citep{shalem2019}.

\begin{figure}
	\centering
	\includegraphics[width=6 cm]{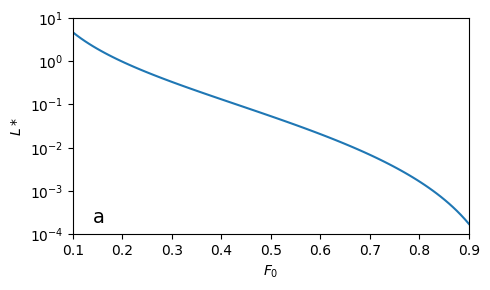}
	\hfill
	\includegraphics[width=6 cm]{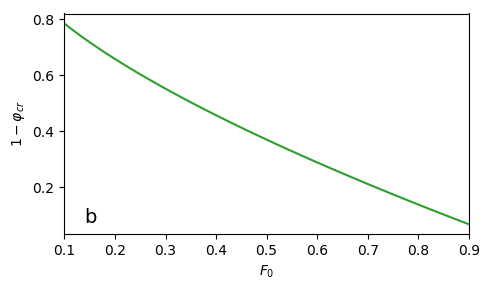}
	\vfill
	\includegraphics[width=6 cm]{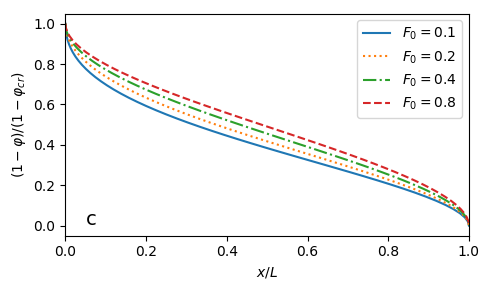}
	\hfill
	\includegraphics[width=6 cm]{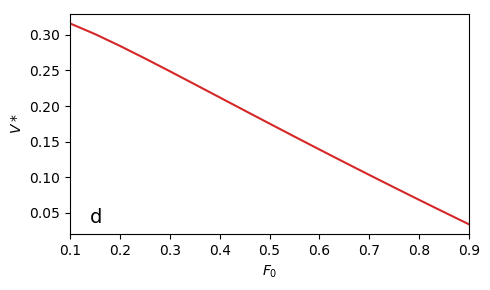}
	\caption{Salt-wedge parameters given as a function of internal Froude numbers: a) non-dimensional salt-wedge length, b) non-dimensional thickness of the salt-wedge at the hydraulic control, c) non-dimensional interface function, d) non-dimensional salt-wedge volume.}
	\label{fig:SW_parameters}
\end{figure}

These non-dimensional salt-wedge parameters are presented in Figure \ref{fig:SW_parameters} in relation to internal Froude numbers. Non-dimensional friction length is shown in Fig.~\ref{fig:SW_parameters}a, critical lower layer depth in Fig.~\ref{fig:SW_parameters}b, salt-wedge interface in Fig.~\ref{fig:SW_parameters}c and seawater volume in Fig.~\ref{fig:SW_parameters}d. Two-layer hydraulic theory predicts that larger values of $F_0$ correspond to shorter salt-wedge intrusion lengths, thinner seawater layers, and lower volumes of seawater in estuaries. The shape of the pycnocline is less sensitive to $F_0$. 

Considering that the internal Froude number is a function of the river inflow and estuary depth, it also follows that salt-wedge intrusion length, thickness and volume increase with the estuary depth and decrease with the river inflow, which has been confirmed by field measurements in microtidal estuaries \citep{krvavica2016field,krvavica2016salt}. Since SLR directly increases the estuary depth, it should also increase the salt-wedge intrusion length and volume.
Note, that the salt-wedge intrusion in a single estuary may shift tens of kilometres upstream or downstream depending on river inflows and sea levels \citep{geyer2011}.

\subsection{Hydrodynamic model for two-layer flow in salt-wedge estuaries with irregular geometry}

\subsubsection{Governing equations}

A numerical model STREAM-1D (STRratified EstuArine Model) \citep{krvavica2016salt,krvavica2017} is similarly based on one-dimensional time-dependent two-layer shallow water equations. It shares the simplifications and assumptions with the two-layer theory defined by Eqs.~(\ref{eq:CMass1})-(\ref{eq:CMom2}). The governing equations, however, are extended to account for irregular channel geometry, as well as a full shear stress and entrainment parametrization. Therefore, STREAM-1D may be used to simulate the dynamics of salt-wedges in channelized estuaries under different combinations of sea levels and river flows rates.

The governing equations are written as the following coupled system of conservation laws with source terms \citep{krvavica2016salt}:
\begin{equation}
\frac{\partial \textbf{w}}{\partial t} +
\frac{\textbf{f}(\boldsymbol{\sigma},\textbf{w})}{\partial x} = 
\textbf{B}(\boldsymbol{\sigma},\textbf{w})\frac{\partial \textbf{w}}{\partial x} +
\textbf{v}(\boldsymbol{\sigma},\textbf{w}) +
\textbf{g}(\boldsymbol{\sigma},\textbf{w}) +
\textbf{s}(\boldsymbol{\sigma},\textbf{w}),
\label{eq:consys1}
\end{equation}
where $\textbf{w}$ is the vector of conserved variables:
\begin{equation}
\textbf{w} = 
\begin{Bmatrix}
A_{1} &  Q_{1} & A_{2} & Q_{2}
\end{Bmatrix}^\text{T},
\end{equation}
vector $\textbf{f}(\boldsymbol{\sigma},\textbf{w})$ is the flux:
\begin{equation}
\textbf{f}(\boldsymbol{\sigma},\textbf{w}) = 
\begin{Bmatrix}
Q_{1} &
\tfrac{Q^{2}_{1}}{A_{1}} + \tfrac{g}{2\sigma_{1}}A^{2}_{1} &
Q_{2} &
\tfrac{Q^{2}_{2}}{A_{2}} + \tfrac{g}{2\sigma_{2}}A^{2}_{2}
\end{Bmatrix}^\text{T},
\end{equation}
and term $\textbf{B}(\boldsymbol{\sigma},\textbf{w})\frac{\partial \textbf{w}}{\partial x} $ appears as a result of coupling the two-layer system, where matrix $\textbf{B}(\boldsymbol{\sigma},\textbf{w})$ is defined as
\begin{equation}
\textbf{B}(\boldsymbol{\sigma},\textbf{w}) = 
\begin{bmatrix}
0 & 0 & 0 & 0 \\
0 & 0 & -g\tfrac{A_{1}}{\sigma_{1}} & 0 \\
0 & 0 & 0 & 0 \\
-gr\tfrac{A_{2}}{\sigma_{1}} & 0 & 0 & 0
\end{bmatrix}.
\end{equation}
where $A_i$ is the layer cross section area, $Q_i$ is the layer flow rate, and  $\boldsymbol{\sigma} = \left\lbrace \sigma_1 \enspace \sigma_2 \enspace \sigma_3 \right\rbrace ^T$ is the vector of channel widths.

The additional source terms have been split into three parts; the first vector $\textbf{v}(\boldsymbol{\sigma},\textbf{w})$ corresponds to derivatives of $\boldsymbol{\sigma}$:
\begin{equation}
\textbf{v}(\boldsymbol{\sigma},\textbf{w}) = 
\begin{Bmatrix}
0 &
\tfrac{g}{2}\frac{\partial}{\partial x}\left(\tfrac{1}{\sigma_{1}} \right)A^{2}_{1} &
0 &
\tfrac{g}{2}\frac{\partial}{\partial x}\left(\tfrac{1}{\sigma_{2}} \right)A^{2}_{2}
\end{Bmatrix}^\text{T},
\end{equation}
the second vector $\textbf{g}(\boldsymbol{\sigma},\textbf{w})$ corresponds to the channel bed, width, and wetted cross-section area (\textit{i.e.}, irregular geometry):
\begin{equation}
\textbf{g}(\boldsymbol{\sigma},\textbf{w}) = 
\begin{Bmatrix}
0 \\
gA_{1}\left[\frac{1}{\sigma_{1}}\frac{\partial}{\partial x}(A_{1} + A_{2}) -
\frac{\partial}{\partial x}(b + h_{2} + h_{1}) \right] \\
0 \\
gA_{2}\left[\frac{1}{\sigma_{2}}\frac{\partial A_{2}}{\partial x} +
\frac{r}{\sigma_{1}}\frac{\partial A_{1}}{\partial x} -
\frac{\partial}{\partial x}(b + h_{2} + rh_{1}) \right] \\
\end{Bmatrix}.
\label{eq:sg}
\end{equation}
and the third vector $\textbf{s}(\boldsymbol{\sigma},\textbf{w})$ corresponds to the friction and entrainment:
\begin{equation}
\textbf{s}(\boldsymbol{\sigma},\textbf{w}) = \textbf{s}_\textrm{F}(\boldsymbol{\sigma},\textbf{w}) + \textbf{s}_\textrm{E}(\boldsymbol{\sigma},\textbf{w}),
\label{eq:sourceterm}
\end{equation}
\begin{equation}
\textbf{s}_\textrm{F}(\boldsymbol{\sigma},\textbf{w}) = 
\begin{Bmatrix}
0 \\
\frac{\tau_{w}}{\rho_1}P_1 + \frac{\tau_i}{\rho_1}\sigma_3\\
0 \\
\frac{\tau_{b}}{\rho_2}P_2 - \frac{\tau_i}{\rho_2}\sigma_3\\
\end{Bmatrix},
\quad
\textbf{s}_\textrm{E}(\boldsymbol{\sigma},\textbf{w}) = 
\begin{Bmatrix}
\frac{1}{r}w_e\sigma_3 \\
\frac{Q_1}{A_1}w_e\sigma_3 \\
- w_e\sigma_3 \\
- \frac{Q_2}{A_2}w_e\sigma_3 \\
\end{Bmatrix},
\label{eq:s}
\end{equation}
where $\tau_w$ is shear stress between upper layer fluid and channel walls, $\tau_i$ is the shear stress at the interface between the layers, $\tau_b$ is the shear stress between the lower layer fluid and channel bed, and $w_e$ is the entrainment velocity.

\subsubsection{Numerical scheme}

The numerical scheme used to solve the governing equations is based on the finite volume method, specifically, the efficient semi-analytical Roe scheme \citep{krvavica2018analytical}. The scheme is explicit in time and stable under the usual CFL condition. It is well balanced, shock-capturing, and considered to be the most accurate first-order scheme. STREAM-1D has been thoroughly tested and validated for other salt-wedge estuaries \citep{krvavica2016salt,krvavica2017}.

%

\section{Results}

The results section consists of two parts. First, we theoretically assess the impacts of SLR on the seawater intrusion in idealized estuaries, and then, we demonstrate the impacts of potential SLR in the Neretva River Estuary in Croatia to validate and illustrate some limitations of theoretical analysis. Furthermore, this practical example is presented to emphasize the importance of a more comprehensive assessment of SLR impacts in salt-wedge estuaries. 

As defined previously, the impacts of SLR on salt-wedge intrusion in ideal estuaries and the Neretva River Estuary are assessed by considering three indicators: 
\begin{itemize}
\item the salt-wedge intrusion length $L$,
\item the salt-wedge volume $V$,
\item the corrective river inflow needed to restore the baseline condition $q_{corr}$. 	
\end{itemize}
The salt-wedge intrusion length is a standard indicator for assessing the impacts of SLR. The salt-wedge volume reflects both the salt-wedge intrusion length and an overall increase of the pycnocline depth. The third indicator is introduced from the viewpoint of managing or mitigating the negative effects of SLR by increasing the river inflow to ensure that sufficient freshwater is available to restore the baseline intrusion length and ensure that sufficient freshwater is available for irrigation or water supply needs. This measure, however, is possible only if some man-made system, such as a hydro-power plat or accumulation, is built upstream from the estuary.

\subsection{Sea-level rise impacts in idealized estuaries}

Sea-level rise in idealized estuaries is accounted for by considering a range of estuary depth increases:
\begin{equation}
H_{slr} = H_0 + \Delta H,
\end{equation}
where $\Delta H$ denotes the water level increase resulting from SLR. To illustrate general SLR impacts, we considered values ranging from $\Delta H=0$ cm (baseline) to $\Delta H=+100$ cm (maximum rise), and estuary channel depths ranging from 3 m to 15 m (which represents the majority of salt-wedge estuaries around the world). Internal Froude number are constrained between $0.1 < F_0 < 0.9$. Note that when $F_0$ approaches one, the flow becomes critical and the salt-wedge is completely pushed out from an estuarine channel. On the other hand, when $F_0$ approaches zero, the intrusion length goes to infinity and an estuary becomes completely occupied by the seawater.

SLR impacts on salt-wedge intrusion lengths are estimated by computing salt-wedge lengths from Eq.~(\ref{eq:L}) for a given sea-level rise $L_{slr}=L(H_{slr}, F_{slr})$ and then comparing them to baseline lengths $L_0=L(H_0, F_0)$, where the internal Froude number affected by SLR is given by $F_{slr} = F(q, H_{slr}, r)$ and baseline internal Froude number by $F_{0} = F(q, H_{0}, r)$.
Similarly, SLR impacts on the salt-wedge volume are estimated by computing the volume of seawater below the pycnocline from Eq.~(\ref{eq:V}) for a given sea-level rise $V_{slr}=V(L_{slr}, h_{2, slr})$ and comparing them to baseline volumes $V_0=V(L_0, h_2)$.
Furthermore, if we denote the baseline river inflow by $q$, and the corrective river inflow (needed to restore the baseline salt-wedge intrusion) by $q_{corr}$, then it is easy to solve the following equation for $q_{corr}$:
\begin{equation}
L\left(H_0, F(q, H_0, r)\right) - L\left(H_{slr}, F(q_{corr}, H_{slr}, r)\right) = 0.
\end{equation}
To quantify absolute values in addition to relative increases, we assumed relative density $r=0.97$ and interfacial friction factor $C_f=5 \times 10^{-4}$, which is a gross estimate suggested by \cite{geyer2011}.

\subsubsection{Assessing sea-level rise impacts by the salt-wedge length}

Theoretical equation (\ref{eq:L}) for the salt-wedge length indicates that SLR increases the salt-wedge intrusion length from $L_0$ to $L_{slr}$ by reducing the internal Froude number from $F_0$ to $F_{slr}$ and increasing the total estuary depth from $H_0$ to $H_{slr}$. Note that SLR affects each estuary differently, the same absolute increase of SLR results in different relative changes of internal Froude number and channel depth.
In other words, the magnitude of potential SLR impacts will depend on the estuary depth and river inflow. 

\begin{figure}
	\centering
	\includegraphics[width=3.9 cm]{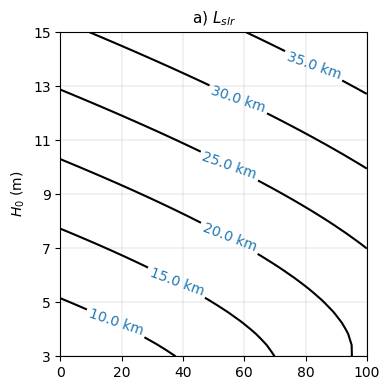}
	\hfill
	\includegraphics[width=3.9 cm]{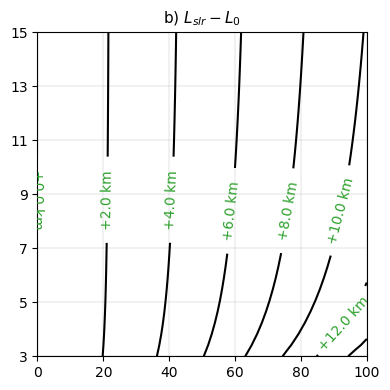}
	\hfill
	\includegraphics[width=3.9 cm]{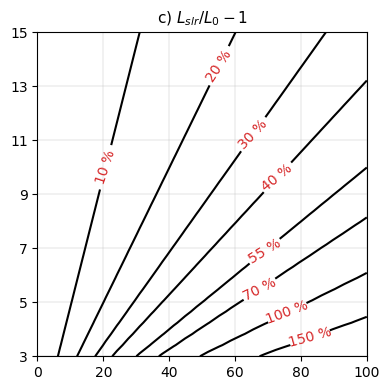}
	\vfill
	\includegraphics[width=3.9 cm]{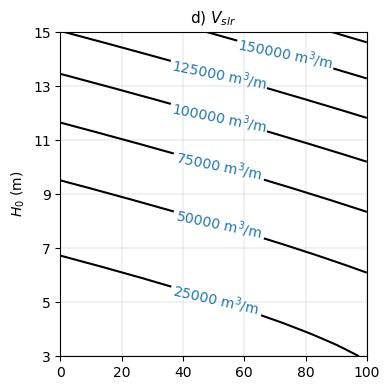}
	\hfill
	\includegraphics[width=3.9 cm]{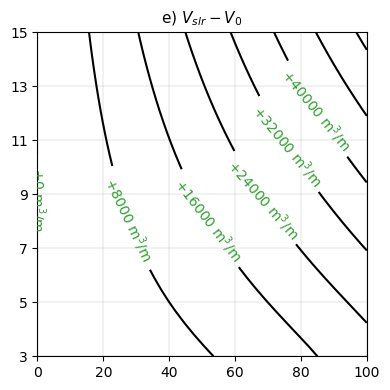}
	\hfill
	\includegraphics[width=3.9 cm]{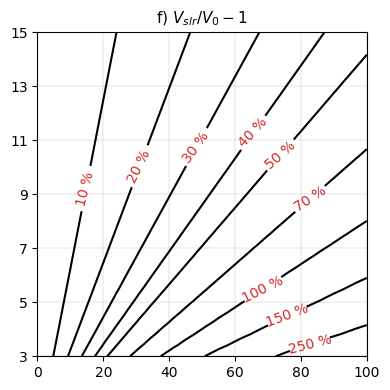}
	\vfill
	\includegraphics[width=3.9 cm]{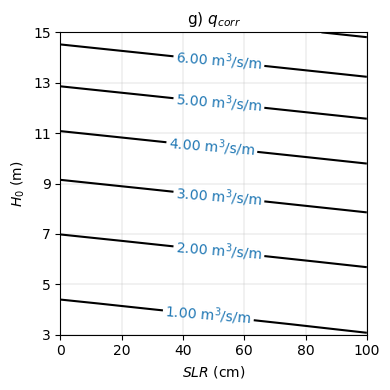}
	\hfill
	\includegraphics[width=3.9 cm]{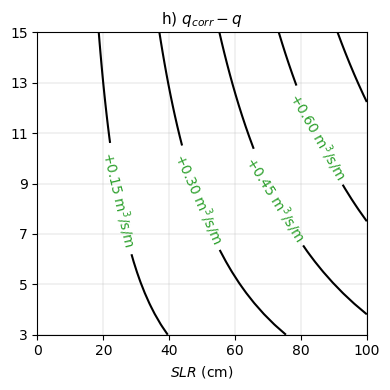}
	\hfill
	\includegraphics[width=3.9 cm]{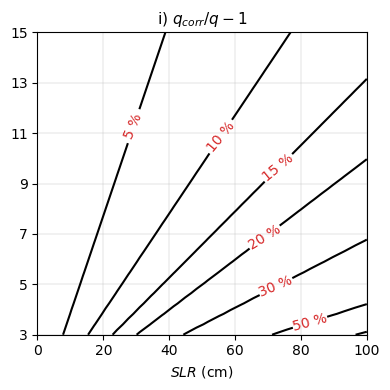}
	\caption{Theoretical impact of SLR on salt-wedge intrusion for $F_0 = 0.2$, with  $r=0.97$ and $C_f=5 \times 10^{-4}$: a) total intrusion length, b) absolute length increase, c) relative length increase, d) total salt-wedge volume, e) absolute volume increase, f) relative volume increase, g) total corrective river inflow, h) absolute inflow increase, i) relative inflow increase}
	\label{fig:theoretical_inc}
\end{figure}

Figs.~\ref{fig:theoretical_inc}a-c shows theoretical increases of salt-wedge intrusion lengths under rising sea-levels, for a predefined range of estuary depths and a low internal Froude number $F_0=0.2$. Similar relative values are obtained for other internal Froude numbers, but, obviously, absolute values vary according to results previously shown in Fig.~\ref{fig:SW_parameters}. As expected, SLR increases the salt-wedge intrusion length under all conditions (Fig.~\ref{fig:theoretical_inc}a). The results show that shallower estuaries are more affected by SLR - salt-wedge intrusion lengths increase more in shallow estuaries for the same SLR (Figs.~\ref{fig:theoretical_inc}b, c). 
A recent study on a large sample of UK estuaries also reported that SLR impacts are more pronounced in shallower ($<10$ m) and micro-tidal estuaries \citep{prandle2015}.


\subsubsection{Assessing sea-level rise impacts by the salt-wedge volume}

In addition to increasing the salt-wedge intrusion length, SLR also rises the average pycnocline depth and consequently the total thickness of the salt-wedge. This dual increase is expressed by the salt-wedge volume defined by Eq.~(\ref{eq:V}). 

Figs.~\ref{fig:theoretical_inc}d-f shows theoretical increases of salt-wedge volumes under rising sea-levels, for a predefined range of estuary depths and a low internal Froude number $F_0=0.2$. Same as for the intrusion length, SLR increases the salt-wedge volume under all conditions (Fig.~\ref{fig:theoretical_inc}d). The results also suggest that shallower estuaries are more affected by SLR. Although absolute increases of the salt-wedge volume are larger in deeper estuaries, the relative increases are much higher in shallower estuaries for the same SLR (Fig.~\ref{fig:theoretical_inc}f). 


\subsubsection{Assessing sea-level rise impacts by the corrective river inflow}

The corrective river inflow defines an additional flow rate needed to push out the salt-wedge to its baseline position. By increasing the river inflow, we increase the internal Froude number. Eq.~(\ref{eq:L}), however, shows that it is not enough to just restore $F_0$ from $F_{slr}$, but the internal Froude number must be larger than $F_0$ to compensate for higher $H_{slr}$.

Figs.~\ref{fig:theoretical_inc}g-i shows theoretical increases of the corrective river inflow under rising sea-levels, for a predefined range of estuary depths and a low internal Froude number $F_0=0.2$. 
We see that for all conditions, the corrective river inflow needs to increase to balance longer salt-wedge intrusion  (Fig.~\ref{fig:theoretical_inc}g). 
The results again suggest that shallower estuaries are more affected by SLR. Similarly as for the salt-wedge volumes, absolute increases are larger in deeper estuaries, but the relative increases are much higher in shallow estuaries for the same SLR (Fig.~\ref{fig:theoretical_inc}i). 


\subsection{Sea-level rise impacts in the Neretva River Estuary}

\subsubsection{Numerical model validation}

Two-layer shallow water models have been previously applied and validated for the Neretva River Estuary \citep{ljubenkov2012} and several other microtidal estuaries \citep{ljubenkov2015,krvavica2016salt}. The numerical results in the Neretva River Estuary showed a good agreement with salinity measurements throughout the year, under a wide range of hydrological conditions \citep{ljubenkov2012}. Furthermore, the STREAM-1D model was thoroughly tested and validated in the Rje\v{c}ina River estuary, which is also located on the Adriatic coast and exposed to identical tidal conditions \citep{krvavica2016salt,krvavica2017}. Additionally, field measurements at several river estuaries on the Adriatic coast \citep{krvavica2016field,ljubenkov2012,ljubenkov2015}, showed that a strong stratification is persistent throughout the year because of microtidal conditions and that the two-layer theory is a reasonable approximation.

In this study, we validated the model after calibrating the channel bed roughness and interfacial friction factor in the Neretva River Estuary and comparing intrusion lengths, as well as upstream water levels to measurements. The validation is based on measurements from 2018 at the river mouth and upstream hydrological station Metkovi\'{c} located 20.9 km from the river mouth (Fig.~\ref{fig:Neretva_geom}b). 

First, we analysed the river inflow and compared the computed salt-wedge length to the measurements of the conductivity at the channel bottom at station Metkovi\'{c}. These measurements show that the salt-wedge tip is located downstream from the Metkovi\'{c} station for $Q>180$ m$^3$/s. Using this constraint, we calibrated the interfacial friction factor and found a good agreement with the measurements for $C_f=6 \times 10^{-4}$. Note, that $C_f$ may slightly vary along the salt-wedge and in time since its magnitude is some function of internal Reynolds and Froude (or bulk Richardson) number \citep{krvavica2017}. However, for steady flow rates, and mild changes in the channel geometry, it can be considered constant. Similar friction factors were found in other salt-wedge estuaries \citep{geyer2011,ljubenkov2012,krvavica2016field}. 

We also compared the numerical results to measured water levels at station Metkovi\'{c}, and found the best agreement for Manning's bed roughness coefficient $n=0.03$ s/m$^{1/3}$. The comparison of computed and measured water levels at station Metkovi\'{c} for the entire 2018, indicated a root mean square error $RMSE = 12$ cm, and Pearsons correlation coefficient $R = 0.96$.


\subsubsection{Assessing SLR impacts on salt-wedge intrusion in the Neretva estuary}

Parameters used for numerical simulations are given by the density ratio $r=0.975$, interfacial friction factor $C_f=6 \times 10^{-4}$, and Manning's bed roughness coefficient $n=0.03$ s/m$^{1/3}$. Spatial step was $\Delta x = 10$ m, and a variable time step was computed based on the stability condition $CFL = 0.8$. The simulation was performed until steady-state conditions were obtained.

First, we performed a series of simulations for three SLR cases, namely 0.0, 0.5 and 1.0 m, and river inflows ranging from 50 to 1000 m$^3$/s, which roughly correspond to the range $F_0 = 0.15 - 0.75$. For each combination of SLR and $Q$ we computed the salt-wedge intrusion length in the Neretva River Estuary. 

\begin{figure}
	\centering
	\includegraphics[width=6 cm]{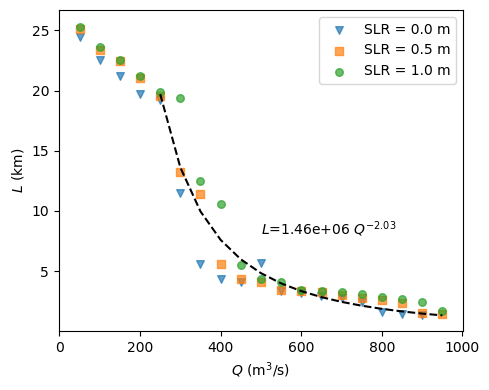}
	\caption{Salt-wedge intrusion lengths for a range of river inflows and three different sea-level rises (R$^2=0.91$)}
	\label{fig:Neretva_QL}
\end{figure}

Cumulative results shown in Fig.~\ref{fig:Neretva_QL} indicate that the intrusion length may be related to the river inflow by the following power law (with R$^2=0.91$):
\begin{equation}
L = 1.46 \times 10^6 \times Q^{-2.03}.
\label{eq:LQ_Neretva}
\end{equation}
Note that the exponent -2.03 is valid for all SLR, but only for higher river inflows, namely $Q>250$ m$^3$/s. At lower river inflows, the channel geometry (steeper channel bed slope) prevents longer intrusion lengths (see Fig.~\ref{fig:Neretva_geom}b). Such exponent agrees with a theoretical range of -2.0 to -2.5 found by \cite{geyer2011} and confirms the importance of the river inflow in stratified estuaries in comparison to partially or well-mixed estuaries, where the exponent is one order of magnitude lower \citep{monismith2002}. Even in some other salt-wedge estuaries, the exponent is usually lower, e.g. -0.19 in the Merrimack River Estuary \citep{ralston2010}.

Quantifying the SLR impacts on the intrusion in the Neretva River Estuary is not so trivial. It seems that uneven channel bed plays an important part in the advancement of the salt-wedge when sea level increases. However, some trends are noticeable (Fig.~\ref{fig:Neretva_QL}). Evidently, SLR increases the salt-wedge intrusion length under all hydrological conditions. The strongest impact of SLR on the intrusion length is observed for river inflows in the range $250 < Q < 500$ m$^3$/s, when SLR may increase the intrusion length by 6 to 8 km further upstream (Fig.~\ref{fig:Neretva_QL}). As stated earlier, for $Q < 250$ m$^3$/s, the expected salt-wedge length increase is prevented by the channel bed elevations, and for $Q > 500$ m$^3$/s the baseline salt-wedge length is much shorter, so that SLR has a lower overall impact $-$ intrusions are increased by up to 2 km.

For a more detailed analysis of the SLR impacts in the Neretva River Estuary, we consider two hydrological scenarios:
Case A defined by the river inflow $Q=300$ m$^3$/s which corresponds to $F_d=0.23$ at the salt-wedge tip, and Case B defined by the river inflow $Q=50$ m$^3$/s which corresponds to $F_d=0.16$ at the salt-wedge tip. Those are roughly the respective mean annual and minimal summer flow rates when the salt-wedge intrusion is most critical. Similarly as in the previous section for the theoretical analysis in idealized estuaries, we considered SLR values ranging from $\Delta H=0$ cm (baseline) to $\Delta H=+100$ cm (maximum rise).

\begin{figure}
	\centering
	\includegraphics[width=6 cm]{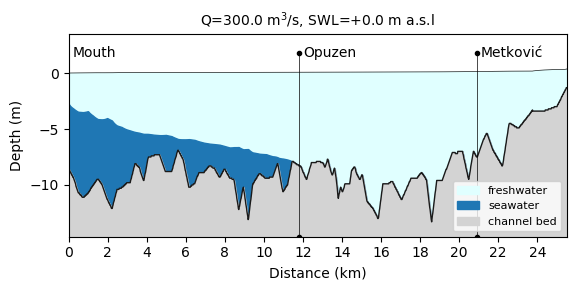}
	\hfill
	\includegraphics[width=6 cm]{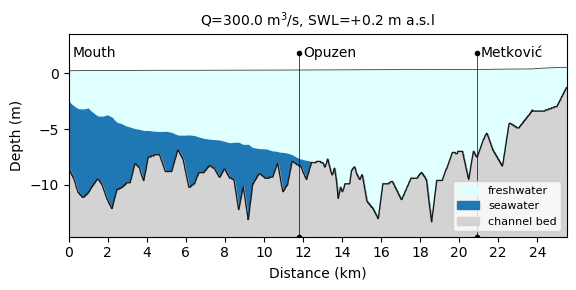}
	\vfill
	\includegraphics[width=6 cm]{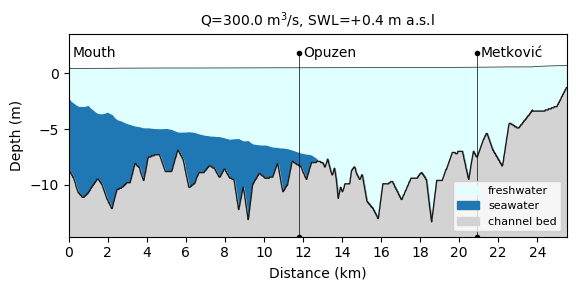}
	\hfill
	\includegraphics[width=6 cm]{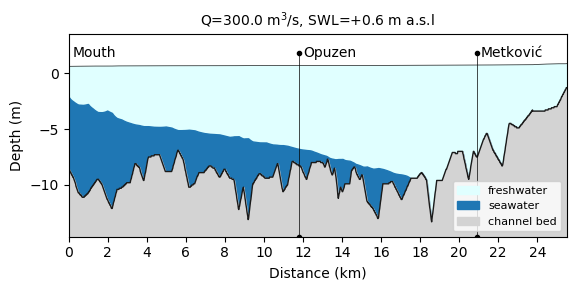}
	\vfill
	\includegraphics[width=6 cm]{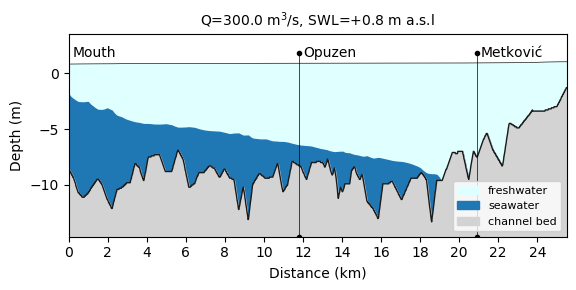}
	\hfill
	\includegraphics[width=6 cm]{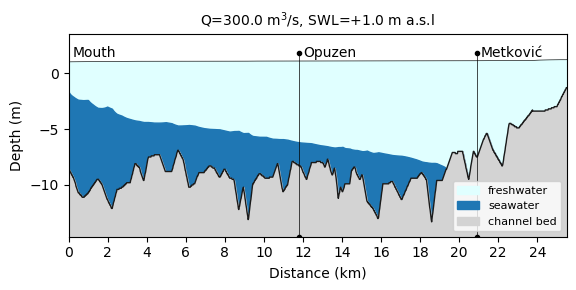}
	\caption{Case A: Longitudinal profiles of computed arrested salt-wedge in the Neretva River Estuary at $Q=300$ m$^3$/s and for $SWL$ ranging from 0 to 100 cm}
	\label{fig:Neretva_Q300_sim}
\end{figure}

Figure \ref{fig:Neretva_Q300_sim} shows the computed arrested salt-wedge profiles for case A, defined by $Q=300$ m$^3$/s and a range of SWL that accounts for SLR. Results show how the salt-wedge intrusion length increases with SLR, starting with $L=11.45$ km for $SWL = +0.0$ m a.s.l., which gradually increases to $L=19.33$ km for $SWL = +1.0$ m a.s.l. We can notice that the increase of the salt-wedge length is not linear because of variable bottom topography which somewhat governs the intrusion, but the changes in intrusion lengths are evident. Total salt-wedge volumes increase from $V=42604$ m$^3$ to $V=77747$ m$^3$. Also, the river inflow needs to increase up to $Q=375$ m$^3$/s to restore the baseline intrusion.

\begin{figure}
	\centering
	\includegraphics[width=6 cm]{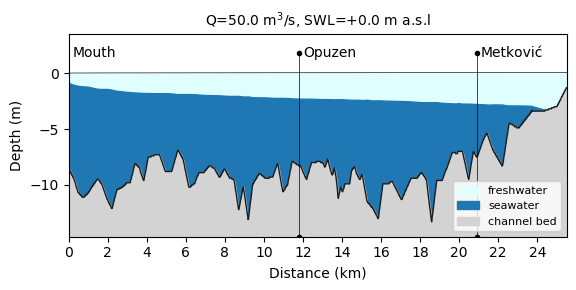}
	\hfill
	\includegraphics[width=6 cm]{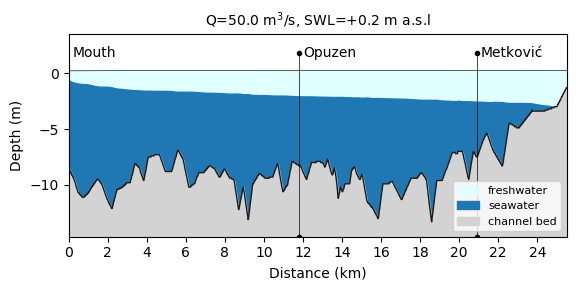}
	\vfill
	\includegraphics[width=6 cm]{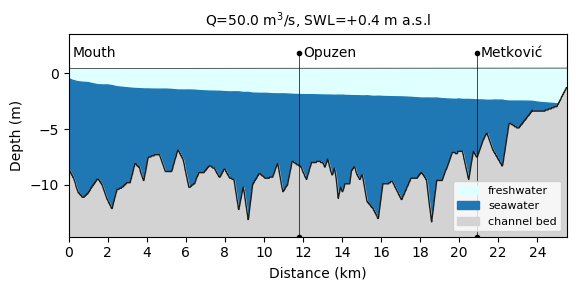}
	\hfill
	\includegraphics[width=6 cm]{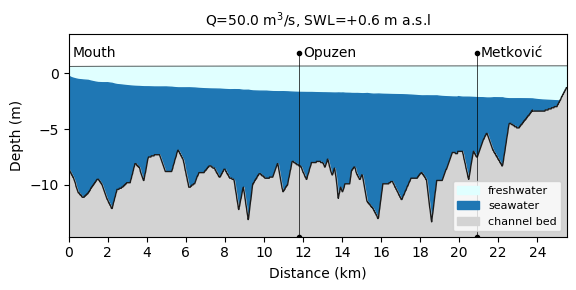}
	\vfill
	\includegraphics[width=6 cm]{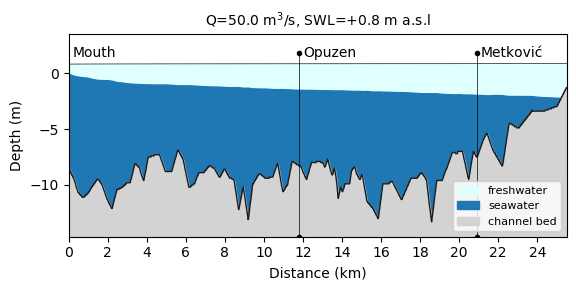}
	\hfill
	\includegraphics[width=6 cm]{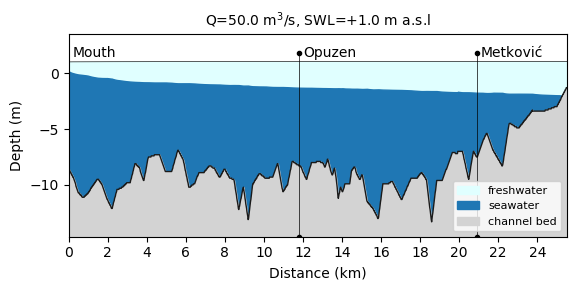}
	\caption{Case B: Longitudinal profiles of computed arrested salt-wedge in the Neretva River Estuary at $Q=50$ m$^3$/s and for $SWL$ ranging from 0 to 100 cm}
	\label{fig:Neretva_Q50_sim}
\end{figure}

Figure \ref{fig:Neretva_Q50_sim} shows the computed arrested salt-wedge profiles for case B, defined by $Q=50$ m$^3$/s and a range of SWL that account for SLR. In contrast to case A, these results show negligible increases in the salt-wedge length. The potential salt-wedge intrusion further upstream is prevented here by a steep channel bed slope. However, an increase in the salt-wedge thickness and volume is evident. Lower layer thickness increases with SLR almost linearly, at Metkovi\'{c} station it changes from $h_2=4.73$ m for baseline SWL to $h_2=5.74$ m for SLR of 1.0 m. Similarly, total salt-wedge volumes increase from $V=160945$ m$^3$ to $V=186054$ m$^3$, and the river inflow needs to increase up to $Q=85$ m$^3$/s to restore the baseline intrusion.

Similar behaviour was found in other estuaries.
\cite{hong2012responses} also found that the pycnocline depth and seawater volume increases with SLR in the Chesapeake Bay.
And \cite{chua2014impacts} also considered that an increase of the river inflow needed to maintain intrusion lengths at predefined locations is a good measure of SLR impacts in San Francisco Bay.

\section{Discussion}

\subsection{How suitable is the theoretical approach for the Neretva River Estuary?}

\begin{figure}
	\centering
	\footnotesize
	Case A: $Q=300$ m$^3$/s ($F_0=0.23$)
	\vfill
	\includegraphics[width=9 cm]{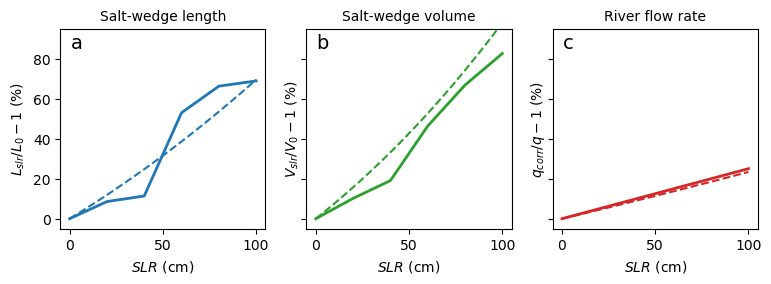}
	\vfill
	\footnotesize
	Case B: $Q=50$ m$^3$/s ($F_0=0.16$)
	\vfill
	\includegraphics[width=9 cm]{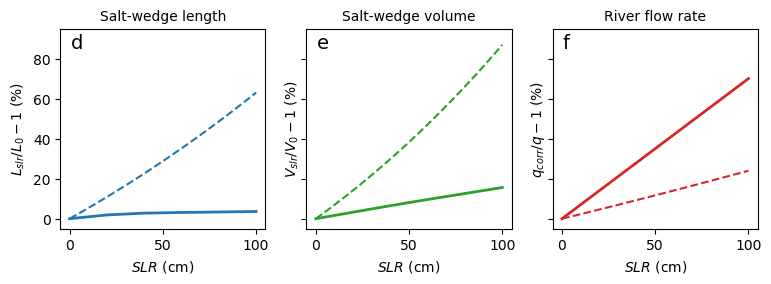}
	\caption{Relative increases of the SLR indicators for cases A and B, where full line ($-$) denotes modelled results for the Neretva River Estuary and dashed line (- -) denotes theoretical results for an idealized 8.5 m deep estuary. First column shows salt-wedge lengths (a, d), second column shows salt-wedge volumes (b, e), and third column shows flow rates needed to restore the baseline intrusion (c, f)}
	\label{fig:Neretva_rel}
\end{figure}

In the previous section, we have presented the potential SLR impacts for idealized estuaries using a theoretical approach and for the Neretva River Estuary using a one-dimensional time-dependent numerical model. This section provides a deeper insight into the performance of proposed simple expressions when applied to a realistic microtidal estuary. For this purpose, we computed the relative changes of the salt-wedge length, volume, and corrective river inflow in the Neretva River Estuary and compared them to theoretical values for an idealized 8.5 m deep estuary. This depth corresponds to the mean value of the Neretva River Estuary bottom function downstream from Metkovic station. Note that the results would differ if some other depth was chosen - shorter salt-wedge lengths would be predicted for shallower depths, an vice versa.
Fig.~\ref{fig:Neretva_rel}a-c shows relative changes in the Neretva River Estuary for case A and Fig.~\ref{fig:Neretva_rel}d-f for case B. 

In the first scenarios (case A) defined by $Q=300$ m$^3$/s and $F_0=0.23$ we notice a very good agreement between theoretical results and the Neretva River Estuary. Relative increases of the salt-wedge length with SLR show some non-linearity, but the overall trend corresponds to theoretical values - for 1.0 m SLR the salt-wedge intrusion length will increase by $69\%$ (Fig.~\ref{fig:Neretva_rel}a). Similar results are obtained for the salt-wedge volume, which is expected to increase by up to $83\%$ for maximum considered SLR (Fig.~\ref{fig:Neretva_rel}b). The river inflow increase needed to restore baseline intrusion in the Neretva River Estuary is almost identical to theoretical values and amounts to $25\%$ for maximum considered SLR.

On the other hand, for the second scenario (case B) defined by a lower river inflow $Q=50$ m$^3$/s and $F_0=0.16$ we notice significant discrepancies between theoretical values and computed changes in the Neretva River Estuary. Salt-wedge intrusion length seems to nearly unaffected by SLR (Fig.~\ref{fig:Neretva_rel}d), although theoretical values in an idealized estuary predict an increase over $60\%$ for 1.0 m SLR. Salt-wedge volumes are increasing with SLR in the Neretva River Estuary (Fig.~\ref{fig:Neretva_rel}e), but at a much lower rate than expected for idealized estuaries. On the other hand, corrective river inflows in the Neretva River Estuary are expected to increase by twice as much than for an idealized estuary (Fig.~\ref{fig:Neretva_rel}f).

The main reason for the discrepancies in the second scenario between theoretical values and numerical results in the Neretva River Estuary is the channel geometry (see Fig.~\ref{fig:Neretva_Q50_sim}). Theoretical salt-wedge intrusion further upstream as predicted in an idealized estuary is prevented by the channel bed in the Neretva River Estuary. Therefore, the salt-wedge length increases only marginally, and the salt-wedge volume primarily reflects a rising pycnocline depth. 
Higher river inflows are also the results of the steeper bed slope - if the salt-wedge is obstructed by the channel bottom, then much higher river inflow is needed to restore its baseline intrusion length.

The first case suggests that small variations over the mean depth have a minor effect on the salt-wedge intrusion. The second case, on the other hand, indicates that steep channel bed slope has a noticeable impact on the seawater intrusion. This is all in agreement with findings by \cite{poggioli2015}, who assessed the sensitivity of salt-wedge estuaries to channel geometry. \cite{yang2015estuarine} also found that the salinity intrusion increases with SLR in the Snohomish River Estuary in the USA are strongly non-linear because of the channel geometry.

\subsection{Extending the theoretical solution for sloped estuaries}

Considering that the channel geometry seems to significantly affect the potential intrusion of a salt-wedge, and that the theoretical approach derived for idealized flat estuaries cannot capture the impact of geometry, we propose an extension of the original ODE that accounts for a channel bed slope. For this purpose, we rewrite Eq.~(\ref{eq:ode}) by introducing a non-uniform channel bed defined by a linear function $b(x) = S_0x$:
\begin{equation}
\frac{\textrm{d}h_1}{\textrm{d}x} =
\frac{F_d^2}{1 - F_d^2} C_f \left(1  +r \frac{h_1}{H_0 - S_0x - h_1}\right).
\label{eq:ode_sloped}
\end{equation}
where $S_0$ is the channel bed slope.

We can introduce the same non-dimensional parameters as before (see Eq.~(\ref{eq:HX})) and obtain
\begin{equation}
\frac{\textrm{d}\varphi}{\textrm{d}\chi}=
\frac{F_0^2}{\varphi^3-F_0^2} \left(1 + r\frac{\varphi}{1-\chi S_0/C_f - \varphi}\right).
\label{eq:odenondim_sloped}
\end{equation}
This ODE, unfortunately, is non-linear for both unknown parameters ($\varphi$ and $\chi$) and, therefore, it is not possible to find an analytical solution, as was the case for a flat estuary. However, Eqs.~(\ref{eq:odenondim_sloped}) or (\ref{eq:ode_sloped}) can still be integrated numerically for some channel slope $S_0$ and interfacial friction factor $C_f$. Notice that the non-dimensional form of governing ODE introduces a new variable $S_0/C_f$ which can be considered a non-dimensional channel bed slope.  

\begin{figure}
	\centering
	\includegraphics[width=6 cm]{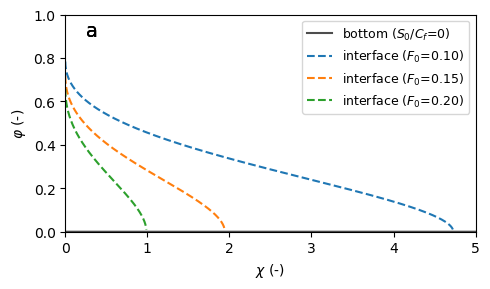}
	\vfill
	\includegraphics[width=6 cm]{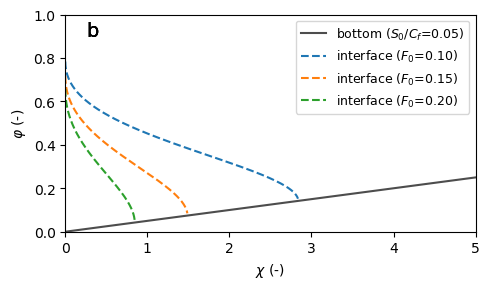}
	\vfill
	\includegraphics[width=6 cm]{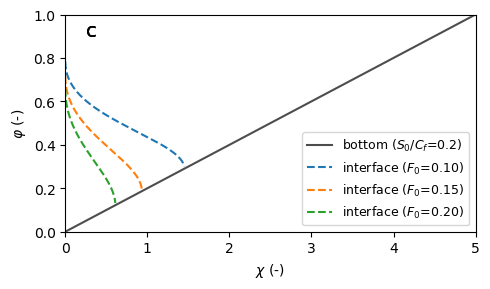}
	\caption{Solution to Eq.~(\ref{eq:odenondim_sloped}) - non-dimensional salt-wedge shape - for internal Froude numbers $F_0=0.1, 0.15, 0.2$ and for: a) horizontal channel bed ($L^*$ equivalent to Eq.~(\ref{eq:LH}), b) mild channel bed slope $S_0/C_f=0.05$, c) steep channel bed slope $S_0/C_f = 0.2$}
	\label{fig:sloped_channels}
\end{figure}

The numerical solutions to Eq.~(\ref{eq:odenondim_sloped}) for three different internal Froude numbers are shown in Fig.~\ref{fig:sloped_channels} for horizontal channel bed, as well as a mild and a steep bed slope. 
The results show that even a mild bed slope can noticeably reduce the potential salt-wedge intrusion length for the same $F_0$ (Fig.~\ref{fig:sloped_channels}b). Figure~\ref{fig:sloped_channels} also indicates that the bed slope impact is more noticeable for lower values of $F_0$. Furthermore, notice that when the channel slope becomes steeper, the distinction between salt-wedged for different $F_0$ is less apparent (Fig.~\ref{fig:sloped_channels}c). This confirms our assumption that the potential salt-wedge intrusion with rising sea-level in the Neretva River Estuary is in fact prevented by a channel bed slope (notice the bed elevations from 18 to 25 km in Fig.~\ref{fig:Neretva_geom}b).

\begin{figure}
	\centering
	\includegraphics[width=6 cm]{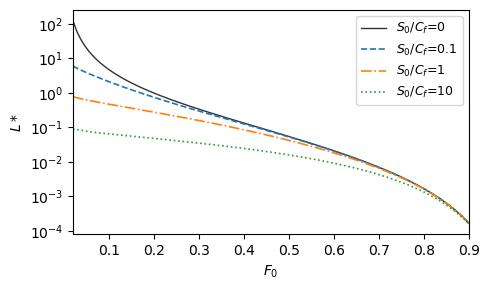}
	\caption{Non-dimensional salt-wedge intrusion length $L^*$ as a function of internal Froude numbers $F_0$ for different non-dimensional channel bed slopes $S_0/C_f$}
	\label{fig:L_F0_sloped}
\end{figure}

In contrast to flat estuaries, where a salt-wedge may intrude indefinitely upstream when river flow and internal Froude number approach zero, in sloped estuaries, channel bed limits the intrusion. Therefore, the non-dimensional salt-wedge intrusion length function branches into multiple curves depending on the non-dimensional slope $S_0/C_f$ as illustrated in Fig.~\ref{fig:L_F0_sloped}. Equation (\ref{eq:odenondim_sloped}) and Fig.~\ref{fig:L_F0_sloped} show that as $Q$ and $F_0$ approach zero, the salt-wedge length approaches a theoretical limit of $C_f/S_0$. This is also in agreement with a theoretical analysis of the salt-wedge sensitivity to channel geometry by \cite{poggioli2015}. 

\subsection{Advantages and limitations of proposed approaches}

Three approaches for assessing SLR impact on salt-wedge intrusion are considered here: a) analytical expressions for idealized estuaries derived from the two-layer theory, b) ODE for sloped estuaries, also derived from the two-layer theory, and c) time-dependant two-layer SWE model for irregular channels derived by extending the two-layer theory. Each of these approaches is increasingly more complex and computationally demanding than the previous one, but with less limitations, and with better accuracy and reliability.

The main limitations of analytical expressions follow from the assumption and simplifications made in deriving them. Temporal effects, such as tidal dynamics, are not accounted for. All contributions from the channel geometry are neglected, including the bed elevations, width variations, as well as transversal and curvilinear effects. Furthermore, the zero thickness pycnocline is rarely found in nature, although in microtidal conditions its thickness may become less than 0.5 m \citep{ljubenkov2012,krvavica2016field}. This also means that all vertical processes, including turbulent mixing, are unaccounted for. On the other hand, this method is very simple, easy to use, and it can provide a quick solution for a wide range of parameters.

The ODE approach retains most of these limitations, but it accounts for the channel bed slope. Although it is more accurate and reliable for real estuaries in contrast to analytical expressions, it needs some numerical method to obtain a solution.

In comparison to the two theoretical approaches, the main advantage of STREAM-1D numerical model is that it incorporates temporal effects and a realistic channel geometry. However, the model still has some limitations, such as 2D and 3D effects. This includes transversal and curvilinear channel effects and turbulent processes at the interface, which includes mixing, vertical transport between the layers, and variation of pycnocline thickness. These processes, however, are of secondary importance in microtidal conditions.

\section{Conclusion}

The impacts of rising sea levels in salt-wedge estuaries are investigated in idealized estuaries using simple expressions derived from a two-layer hydraulic theory and in the Neretva River Estuary in Croatia using a one-dimensional two-layer time-dependent model. The assessment of the SLR impacts was based on three indicators - the salt-wedge intrusion length, the salt-wedge volume, and the river inflows needed to restore the baseline intrusion. 

The theoretical analysis in idealized estuaries showed that potential SLR increases all three considered indicators. The results also revealed that shallower estuaries are more sensitive to SLR than deeper estuaries. In comparison to baseline conditions, the salt-wedge intrusion length may increase by more than $150\%$, the salt-wedge volume by more than $250\%$, and the river inflows may need to increase by more than $50\%$ to restore the baseline conditions, depending on the estuary depth and the extent of SLR.

The numerical results for the Neretva River Estuary confirmed that potential SLR increases the salt-wedge intrusion. The results, however, indicated that several different indicators should be used when assessing SLR impacts. 
The salt-wedge intrusion length may be unreliable at low flow conditions due to its high sensitivity to channel geometry. In those case, the salt-wedge volume may be a better indicator because it also reflects the rise of the pycnocline depth with SLR, which may have a negative effect on the salinity intrusion into the aquifer and soil. Corrective river inflow is the most reliable indicator and the most practical one from the viewpoint of water management, but relevant only when some hydraulic structure system is already implemented upstream from the estuary and sufficient water quantities are available for mitigating the salt-wedge intrusion.

In conclusion, the proposed theoretical approach for SLR impacts can be considered a simple and practical tool suitable for preliminary assessment of salinity intrusions in salt-wedge estuaries worldwide. However, theoretical increases of intrusion lengths are meaningful only when predicting changes lower than maximum possible intrusion lengths, which can be limited by a channel bed slope. 
A numerical integration of an extended ODE for sloped estuaries is, therefore, a more reliable choice when the channel bed has a pronounced effect, which is usually the case for lower river inflows.

Time-dependant two-layer shallow-water models can accurately capture all relevant consequences of SLR in salt-wedge estuaries and are, therefore, a good alternative to 3D hydrodynamic models.
In the Neretva River Estuary, the proposed numerical model was successfully applied to predict impacts from potential SLR and to define appropriate mitigation measures, such as an increase of river inflows needed to restore pre-existing conditions critical for agricultural needs. Overall, the proposed numerical model can be a valuable addition in coastal zone management, especially when assessing the impacts of SLR or any other man-made changes in salt-wedge estuaries.

\section{Acknowledgement}
Funding: This research was supported by the University of Rijeka, grant number 17.06.2.1.02 “Sea and river interaction in the context of climate change” and UNIRI-TECHNIC-18-54 "Hydrology of Water Resources and Identification of Risks from Floods and Mud Flows in Karst".

\small
\bibliographystyle{elsarticle-harv} 
\bibliography{References}

\end{document}